\documentclass[amsmath, amssymb, superscriptaddress, reprint, twocolumn, colorlinks=true, citecolor=blue, linkcolor=blue, urlcolor=blue]{revtex4-1}

\usepackage{natbib}
\usepackage[utf8]{inputenc}
\usepackage[T1]{fontenc}
\usepackage{graphicx}
\usepackage{grffile}
\usepackage{longtable}
\usepackage{wrapfig}
\usepackage{rotating}
\usepackage[normalem]{ulem}
\usepackage{amsmath}
\usepackage{textcomp}
\usepackage{amssymb}
\usepackage{capt-of}
\usepackage{hyperref}
\usepackage{mathptmx}
\usepackage{upgreek}
\usepackage{float}
\usepackage{bm}
\usepackage{setspace}
\usepackage{booktabs}
\usepackage{subfigure}
\usepackage{url}
\usepackage{hyperref}
\usepackage{txfonts}
\DeclareSymbolFont{letters}{OML}{ztmcm}{m}{it}

\date{\today; Corresponding emails: denis.eremin@ruhr-uni-bochum.de}

\begin{document}


\title{Sheath electron heating in surface wave discharges driven at microwave frequencies}

\author{Denis Eremin$^*$}
\affiliation{Institute of Theoretical Electrical Engineering, Ruhr University Bochum, Universitätsstrasse 150, D-44801 Bochum, Germany}



\author{Andrew T. Powis}
\affiliation{Princeton Plasma Physics Laboratory, Princeton, NJ 08540, United States of America}

\author{Igor D. Kaganovich}
\affiliation{Princeton Plasma Physics Laboratory, Princeton, NJ 08540, United States of America}


\begin{abstract}
Using fully electromagnetic particle-in-cell/Monte Carlo simulations, the electron heating due to interaction with a moving sheath is demonstrated to dominate in surface wave-driven discharges at microwave frequencies and relatively low pressures. Electrons impinging on the rapidly expanding sheath gain energy by repulsion from its strongly negative potential, similarly to the corresponding mechanism in capacitively coupled discharges driven at radio frequencies. This results in generation of energetic electron beams propagating towards the bulk plasma. In contrast to the expectations from previous theoretical studies, the electron heating due to plasma resonance is not observed. 
\end{abstract}

\maketitle

Surface wave (SW)-driven discharges are frequently used for plasma-aided technologies, such as thin film deposition, etching, or cleaning. This can be attributed to the ease of robust sustainment of such discharges over large surface areas, and over a wide range of frequencies and working gas pressures \cite{moisan_1991,sugai_1998}. Microwave frequencies provide a high plasma generation efficiency (plasma density produced per fixed power), and a relatively small sheath voltage, reducing the possible damage of reactor materials caused by impinging ions. For deposition technology, it is often important to maintain a relatively low gas pressure, which, in turn, allows for higher fluxes of material to be deposited on the substrate. When $\omega \gg \nu$ and $\lambda_{\epsilon}\gg L$, with $\omega$ the driving frequency, $\nu$ the electron-neutral collision frequency, $\lambda_\epsilon$ the energy relaxation length, and $L$ the characteristic size of the system, such discharges are operating in the kinetic and nonlocal regime, enabling various electron heating mechanisms other than Ohmic heating, which loses efficiency due to reduced collisionality. Electron heating mechanisms under such conditions have been a topic of discussion in the literature, with the prevailing belief that in overdense (with the bulk plasma frequency substantially exceeding the driving wave frequency $\omega$) plasmas they are dominated by plasma resonance at the location where $\omega=\omega_{pe}$ with $\omega_{pe}$ the electron plasma frequency \cite{aliev_1993,ganachev_2003}. There, the electric field should experience a strong increase due to the conservation of total current \cite{Ginzburg1965}, accompanied by generation of plasma waves propagating in the direction of decreasing plasma density \cite{ganachev_2003}, dissipating their energy to electrons by a collisional or collisionless (e.g., Landau) damping mechanism and producing electron beams flying toward the dielectric boundary, along which the SWs propagate. However, experimental studies have detected beams of energetic electrons moving in the direction of increasing plasma density, toward the plasma bulk, contradicting this prediction \cite{kudela_2000,terebessy_2000,boudreault_2012}.  

There is yet another mechanism of electron heating known to prevail, at low pressures, over the Ohmic heating in capacitively coupled radio-frequency (CCRF) plasma discharges \cite{popov_1984}. In this case, electrons can acquire energy through reflections from the moving sheath potential created by the positive space charge of the sheaths \cite{godyak_1976}. For the overdense plasmas in question, this same mechanism can also contribute, since $\omega_{pe} \gg \omega$, and thus the sheath space charge is modulated in time roughly at the wave frequency. Moreover, since the sheath edge velocity in the first approximation is proportional to the wave frequency, this mechanism will grow in importance with the increase of $\omega$. 

In this Letter, we report the first demonstration of the significant role of sheath oscillations in electron heating in low-pressure SW discharges at microwave frequencies using a self-consistent electromagnetic particle-in-cell code. A direct experimental investigation of electron heating dynamics in such dense plasmas ($n_e \simeq 10^{18}$ m$^{-3}$) is highly challenging due to the small thickness of the region where the electron heating takes place (a fraction of a millimeter). The simulations conducted in this work enable a detailed study of the electron energization mechanisms and explicitly show generation of fast electrons, being energized through interaction with the expanding sheath and then propagating in the direction of the plasma bulk, just as observed in experiments. 

\begin{figure}[t]
\includegraphics[width=4.5cm]{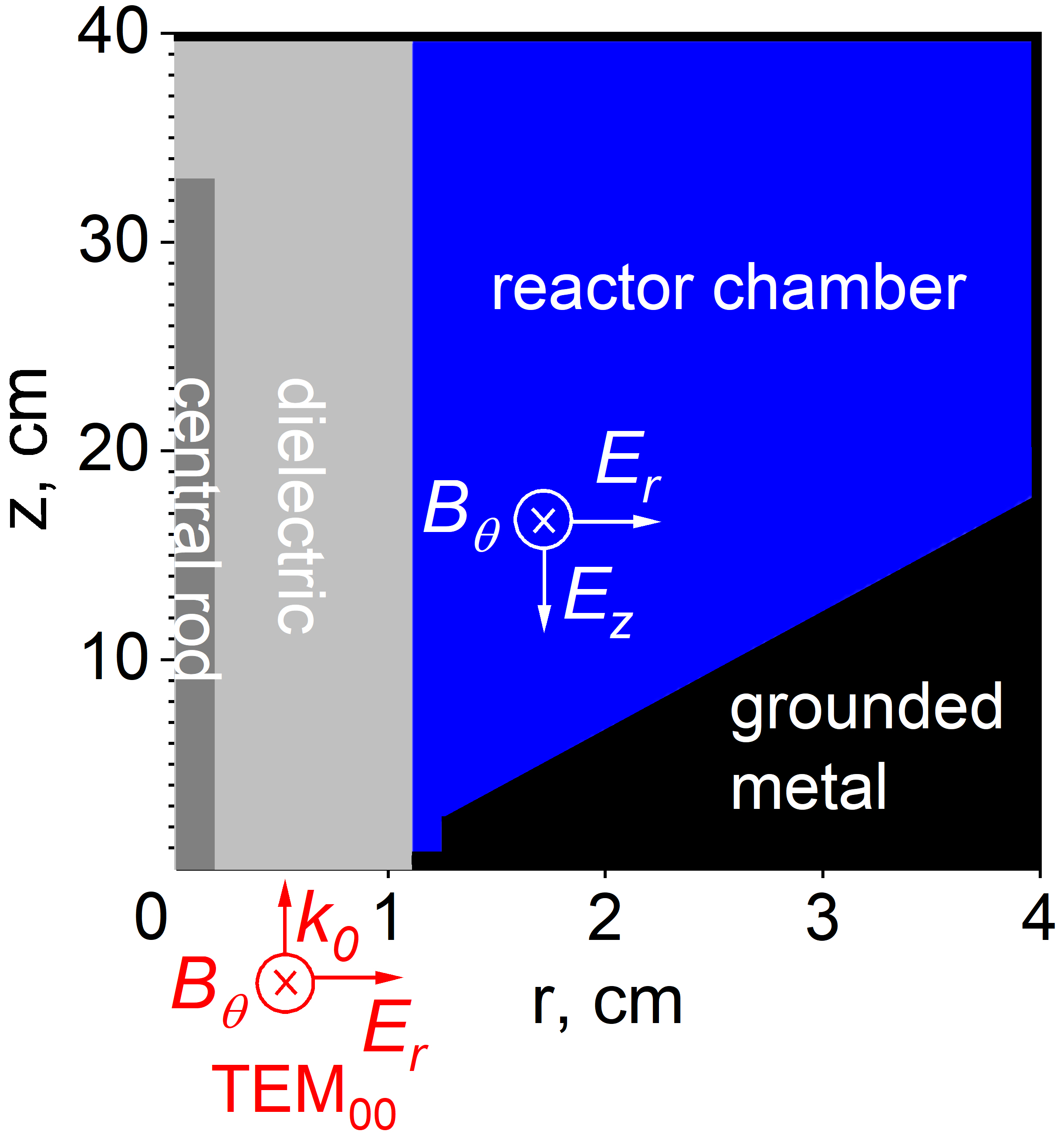}
\caption{\label{fig1} 
Model cylindrical geometry of a MW-driven plasmaline discharge \cite{deilmann_2008,petasch_1997}. The electromagnetic wave enters the plasmaline as a ``coaxial cable'' TEM$_{00}$ mode through the dielectric (light gray) at the bottom ($z=0\,,\, 0.2 \leq r \leq 1.1$ cm), separating the conducting central rod (dark gray) and the grounded conducting reactor wall (black). Plasma is produced in the reactor chamber due to a surface mode propagating further along the dielectric and forming a standing wave upon reflection from the end of the central rod. Note that the plot scales in $r$ and $z$ differ by a factor of 10. 
}
\end{figure}

The model geometry used for the study is shown in Fig.~\ref{fig1}. Although related to the geometry of a specific experimental device used for thin film deposition aimed at reduction of gas permeability for polyethylene terephthalate (PET) bottles \cite{deilmann_2008,deLosArcos_2024}, the configuration has features common to most SW-driven plasma discharges. SWs powering the discharge are generated from a TEM$_{00}$ wave entering the reactor through a dielectric port between the conducting central rod and the grounded metal wall. The TEM$_{00}$ wave is then converted into TM-like SWs propagating in the axial direction along the dielectric surface. Since plasma densities in such reactors are highly overdense, no bulk electromagnetic modes are possible. The SW exists between the central rod and the highly conducting plasma due to the dielectric and the electron-depleted plasma sheath, with the electric field components decaying in the plasma on the skin depth scale. Note that for plasmaline discharges, plasma is created on the outer radius of the dielectric shell \cite{petasch_1997,kossyi_2010}, whereas for surfatrons it is created on the inner radius \cite{moisan_1991} or is produced on a flat surface in planar discharges \cite{ghanashev_2002}. The physics, however, remains similar in all cases.   

\begin{figure}[t]
\centering
\includegraphics[width=8.5cm]{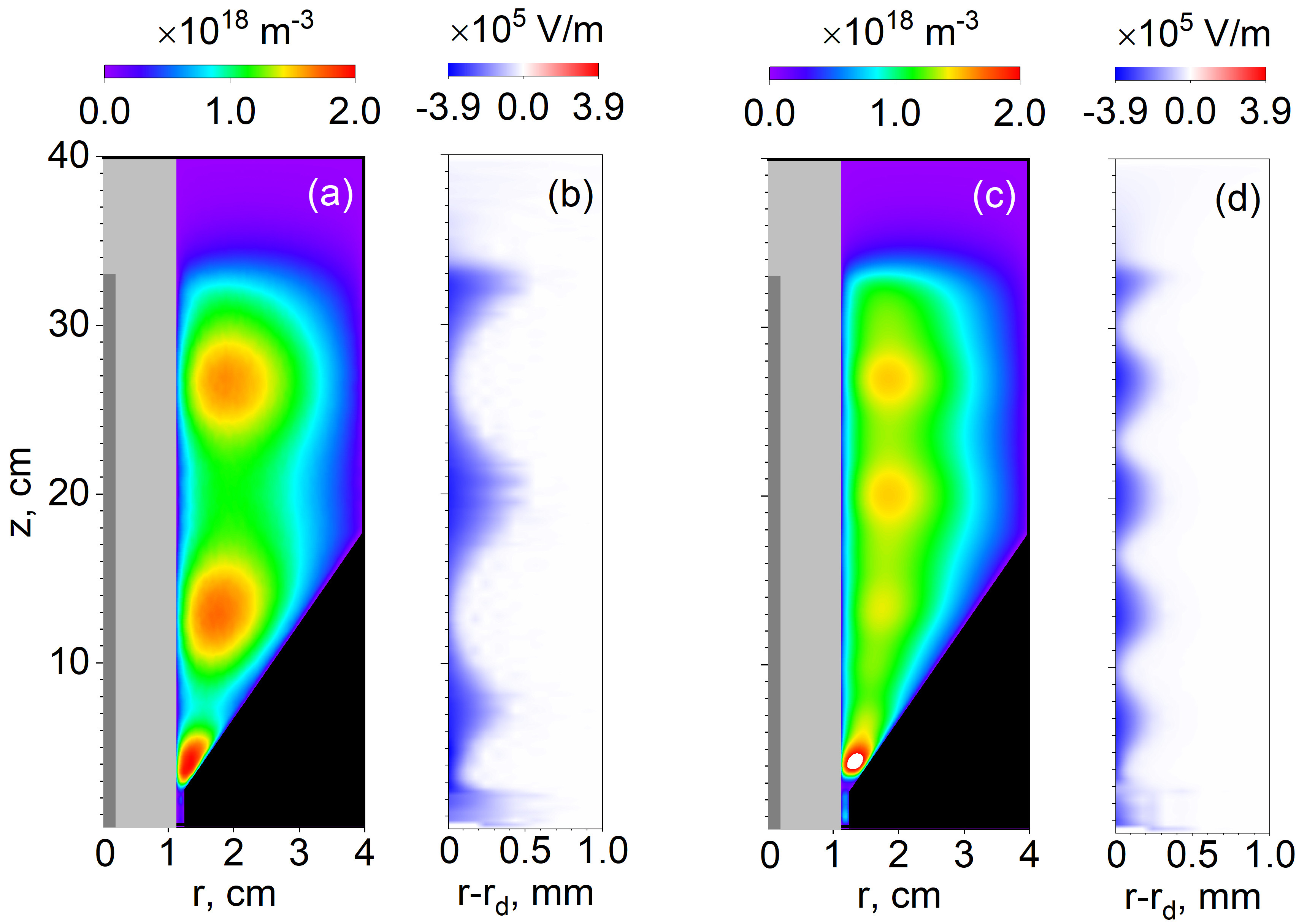}
\caption{\label{fig2} (a) and (c) MW period-averaged electron density for the $500$ MHz and $1$ GHz cases respectively. (b) and (d) MW period-averaged radial electric field
in the vicinity of the dielectric surface $r=r_d$ for the same driving frequencies as in (a) and (c).}
\end{figure}

%

To numerically model the discharge in argon at a relatively low pressure of $10$ Pa, we use the implicit fully electromagnetic charge- and energy-conserving particle-in-cell (PIC) code \nobreak{ECCOPIC2M}, described in detail in \cite{Eremin2023} and validated in \cite{Eremin2023,Berger2023,Eremin2025}. Due to the localization of sheath effects in the vicinity of the dielectric surface, which can be attributed to the large plasma density and relatively small sheath voltage, a strongly nonuniform grid is utilized to resolve this region using the mapped grid algorithm proposed in \cite{chacon_2013}. 

The results demonstrate that excited SWs form standing waves with the number of nodes in the axial profile of the average radial electric field increasing with the driving frequency  (see Fig.\ref{fig2}(b) and \ref{fig2}(d)). However, as the frequency increases from $500$ MHz to $1$ GHz, the peaks of the corresponding plasma density profile (Figs.\ref{fig2}(a) and (b)) change their correlation with the troughs to correlation with the peaks of the average radial electric field (Figs.\ref{fig2}(c) and (d)). The same trend also holds for the axial profile of the ionization rate (not shown), which can be explained by the elongated discharge geometry, leading to much faster plasma diffusion in the radial vs axial direction so that the axial profile of the plasma density is caused by the axial nonuniformity of ionization. Furthermore, other simulations reveal that the correlation exhibited by the $500$ MHz case persists at a lower frequency of $250$ MHz, whereas correlation observed for the $1$ GHz case persisted for a $2.45$ GHz case.

\begin{figure}[t]
\centering
\includegraphics[width=9.0cm]{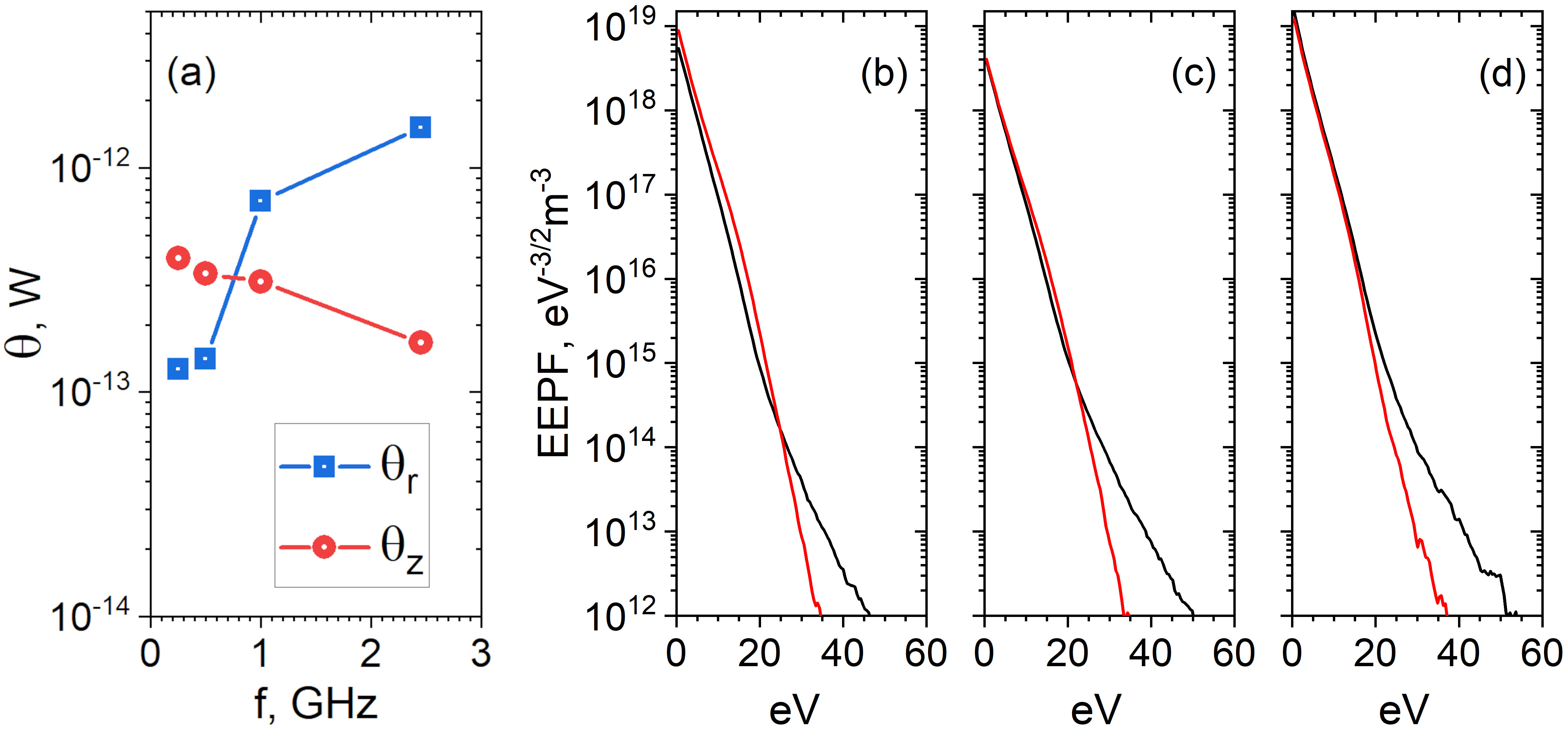}
\caption{\label{fig3} (a) Specific power density $\theta$ (power density absorbed by the plasma divided by the electron density) calculated as $\theta_r = \Delta V^{-1}\int_{\Delta V} d^3r (j_rE_r/n_e)$ and $\theta_z = \Delta V^{-1}\int_{\Delta V} d^3r (j_zE_z/n_e)$ with $\Delta V$ the reactor chamber volume; (b), (c), and (d) electron energy probability functions (EEPFs) provided by PIC simulations in the radial interval of $1$ mm from the dielectric surface at a radial electric field's anti-node (black) or node (red) location for driving frequencies of $500$ MHz, $1$ GHz, and $2.45$ GHz, respectively.}
\end{figure}

This transition can be attributed to a change in the predominant electron heating mechanism. Note that since the standing wave with the nodes (anti-nodes) of the radial electric field coincide with locations of the axial electric field maxima (minima), there can be two primary electron heating mechanisms, Ohmic heating at the nodes, and the virtually collisionless sheath heating at the anti-nodes. Interestingly, this results in the simultaneous presence, at different locations within a single SW discharge, of concave and convex electron energy probability functions (EEPFs) characteristic of the different modes of electron heating observed in CCRF discharges \cite{godyak_1990}(see red and black curves in Fig~\ref{fig3}(b-d)).
At low frequencies (500 MHz and below) electrons above the ionization threshold (in argon $\approx$ 15.8 eV) are produced primarily due to collisional Ohmic heating caused by the axial electric field being strongest between the anti-nodes of the radial electric field, whereas at higher frequencies (1 GHz and above) such electrons are created as a result of virtually collisionless interaction with the moving sheath and the related radial electric field. Indeed, this is further demonstrated by plots of $\theta$, the power density absorbed by electrons divided by electron density, which is indicative of how much power is absorbed on average per electron \cite{moisan_2022}, see Fig.\ref{fig3}(a). Evidently, the radial $\theta$ related to Ohmic heating becomes larger than the axial value related to the sheath heating at $1$ GHz, as expected. Further, comparing EEPFs in the vicinity of the dielectric surface sampled at axial locations of a trough and a peak of the radial electric field (Figs.\ref{fig3}(b)-(d)), one can observe that the electron heating at the troughs is less efficient at producing energetic electrons than at the peaks. Also, although the Ohmic heating generates more energetic electrons 
in total than the sheath heating mechanism, the former appears to be unaffected by the increasing driving frequency, whereas the latter grows with it and becomes dominant everywhere for $2.45$ GHz. In fact, at $1$ GHz it becomes the decisive factor in determining the ionization rate, since the highly energetic electrons have a larger ionization cross-section than those just above the ionization threshold.

\begin{figure}[t]
\includegraphics[width = 8.7cm]{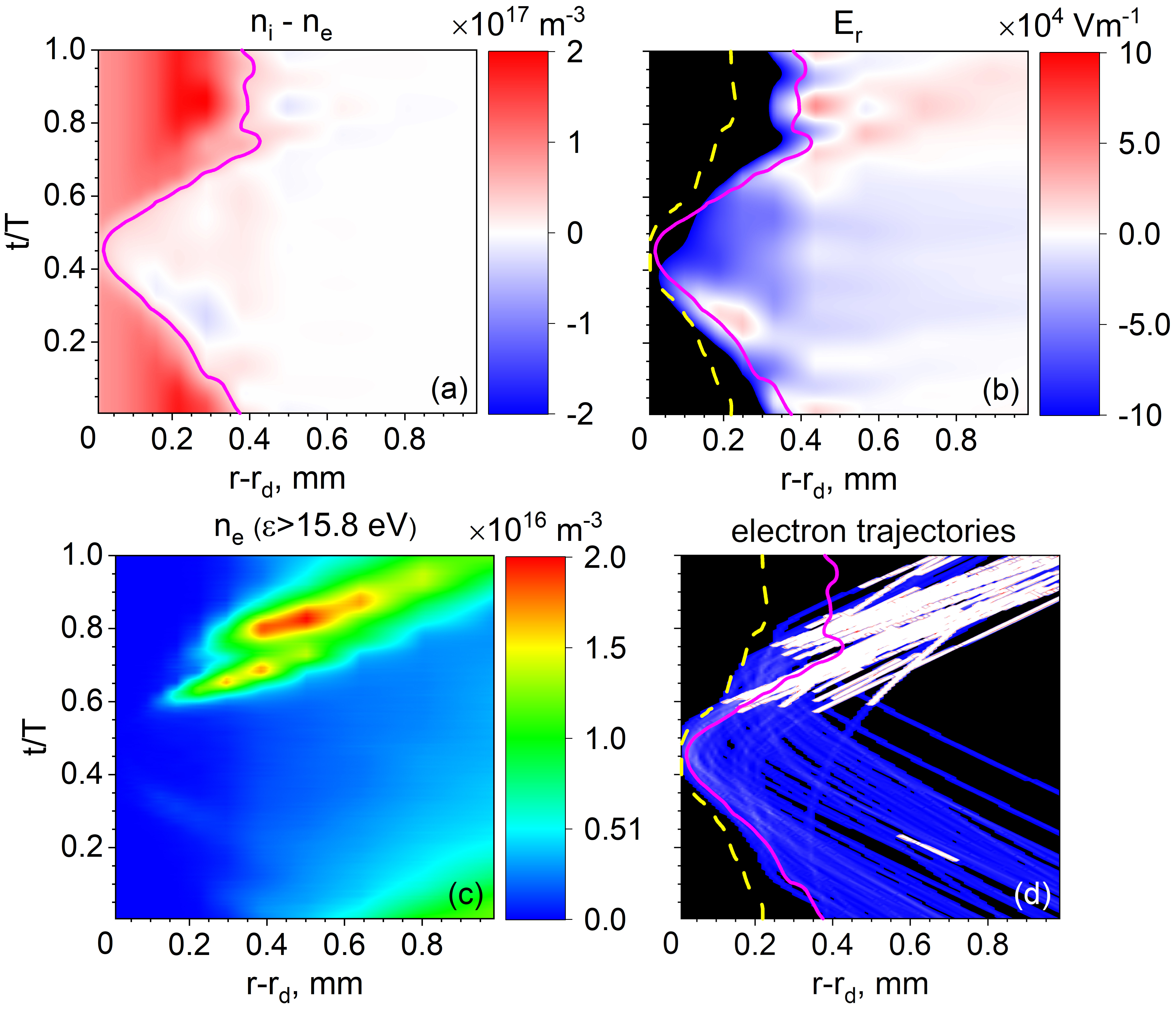}
\caption{\label{fig4} Plots of spatio(radial)-temporal data from the PIC simulation of the $1$ GHz case over one period with the axial location corresponding to that of peak ionization rate: (a) Difference between ion and electron densities and sheath edge location $s(t)$ (solid magenta curve), defined as $n_i(s)-n_e(s) = 0.2 (n_i-n_e)_{max}$. (b) Radial electric field (with cutoff shown in black at a typical value of $-10^5$ V/m experienced by electrons at their reflection from the sheath). (c) Density of electrons with kinetic energy above the ionization threshold for argon. (d) Orbits of some electrons from the PIC simulation. The selection was made based on the criterion that they gained energy above the argon ionization threshold ($15.8$ eV) at the location of the energetic electron beam shown in (c) at $t/T>0.6$, thereby contributing to it. Locations of the sheath edge (solid magenta curve) and the expected plasma resonance (dashed yellow curve) are also shown. The orbit color is painted white whenever kinetic energy of an electron exceeds the ionization threshold. 
}
\end{figure}

The virtually collisionless nature of the electron heating caused by sheath expansion can be directly examined from spatio(radial)-temporal output of the PIC code at a radial electric field anti-node, where such heating is strongest (see Fig.\ref{fig4}). Fig.\ref{fig4}(a) shows the difference between the ion and electron densities, which demonstrates the oscillatory behavior of the plasma sheath at the dielectric surface close to where SWs propagate, Fig.\ref{fig4}(b) shows evolution of the radial electric field close to the dielectric surface, the black area boundary indicating a typical value of the radial electric field experienced by electrons which will go on to form the energetic electron beam. One can see that this boundary closely follows the sheath edge boundary (solid magenta line) deduced from Fig.\ref{fig4}(a). Further, Fig.\ref{fig4}(c) shows the density of the energetic electron beam, and finally, Fig.\ref{fig4}(d) follows orbits of some electrons, which make up the energetic beam, with the sheath edge curve from Fig.\ref{fig4}(a) and the location of plasma resonance (dashed yellow curve) included. Fig.\ref{fig4}(d) proves that the beam electrons can be traced back to electrons coming from the bulk plasma and interacting with the plasma sheath. 

In contrast to the simplified picture, where electrons gain energy from an essentially instantaneous reflection from a moving sheath viewed as a ``hard wall'' \cite{Godyak1972}, in these simulations the electron-sheath interaction is quite intricate and lasts for a substantial fraction of the driving period. In a simplified model, where the radial electric field can be described by $-E_0^\prime(r-s(t))$ (with $s$ the sheath edge location) for $r<s(t)$ and $0$ otherwise, the radial coordinate of an electron in the sheath is governed by the equation of a driven oscillator,
\begin{equation} \label{eq1}
\ddot{\tilde{r}} + \Omega^2\tilde{r} = f,
\end{equation}
where $\tilde{r} = r-s(t)$, $\Omega^2 = eE_0^\prime/m_e$, and $f$ is equal to $-\ddot{s}$ for $\tilde{r}<0$ and $0$ otherwise. The sheath interaction duration can therefore be estimated as $\Delta T_{si}=\pi/\Omega$, which for the $1$ GHz case shown in Fig.\ref{fig4} yields $\Delta T_{si}/T \approx 0.2$ with $E_0^\prime \approx 1.6\times 10^9$ V/m$^2$ as measured from the corresponding simulation. Such an estimate agrees well with the average electron-sheath interaction time that can be evaluated by analyzing the radial electric field experienced by the electrons shown in Fig.\ref{fig4}(d) and the corresponding intervals of large negative electric field. 
There is a large number of such electrons that start interacting with the sheath while it is still undergoing collapse and then remain within the sheath for some time during its expansion.
An analysis of separate electrons orbits reveals that it is also important to take into account their interaction with the pre-sheath radial electric field that remains in the sheath area $r<0.4$ mm even when the sheath edge has retreated. This field is mostly negative (except during local field reversal in the interval $t/T=(0.2,0.4)$ at $r-r_d = 0.2$ mm caused by a double layer in the charged density and resulting in a positive electric field similar to some CCRF discharges \cite{Vender1992,Sato1990}) and is modulated in time only weakly, contrary to the electric field of the sheath proper. The field within the pre-sheath can be attributed to the ambipolar electric field in CCRF discharges \cite{wilczek_2020}. 
It can be approximated that electrons start interacting with the expanding sheath when their velocity is negligible compared to the sheath expansion velocity.
For the case with $1$ GHz shown in Fig.\ref{fig4}(a) one can estimate the sheath edge expansion speed as being nearly constant and equal to $1.4 \times 10^6$ m/s, which yields $m_e(2\dot{s})^2/2 \approx 22$ eV, agreeing well with the average electron energy gain of approximately $25$ eV observed for such electrons in the simulation. Note that Fig.\ref{fig4}(a) shows that sheath expansion occurs substantially faster than sheath retreat. Such an asymmetric behavior can be explained by surface mode excitation, addressed below.



Clearly, electron energization is a complex process \cite{Kaganovich_2002} involving self-consistent dynamics between electrons and the radial electric field in the pre-sheath and sheath regions. Electrons flying in the direction of the sheath interact with the oscillating electric field in the pre-sheath, those that have a sufficient energy to reach the sheath interact with the decelerating ambipolar electric field and accelerating reversed electric field during the sheath collapse, whereupon they are accelerated by the electric field of the moving sheath, which results in the formation of energetic electron beams. The beams, in turn, trigger oscillations in plasma \cite{wilczek_2016,haomin_2022}, which can be observed in Fig.\ref{fig4}(b) starting at $t/T\approx 0.6$. Fig.\ref{fig5} demonstrates the radial phase space dynamics of the electron distribution from $t/T=0.2$ (just before sheath collapse) up to $t/T=0.7$ (peak electron beam generation). At $t/T=0.3$, electrons moving towards the sheath have higher than average energy, apparently facilitated by the positive electric field at $r-r_d = 0.2$ mm. These electrons remain in sync with the sheath and feel its electric field, which increases in time due to sheath expansion. Once electrons detach from the sheath starting at $t/T=0.6$, they have gained sufficient energy to exceed the ionization threshold shown by the dashed white line. At $t/T=0.7$ one can observe onset of the second electron beam formation, which can be ascribed to the interaction between the electric field oscillations excited by the first beam and cold electrons at the sheath edge \cite{wilczek_2016}. An estimate of the parameters for the electron beams observed in Fig.\ref{fig4}(c) places them in the area of the parametric decay instability based on Fig. 2 in \cite{haomin_2022}. 

\begin{figure}[t]
\centering
\includegraphics[width=8.5cm]{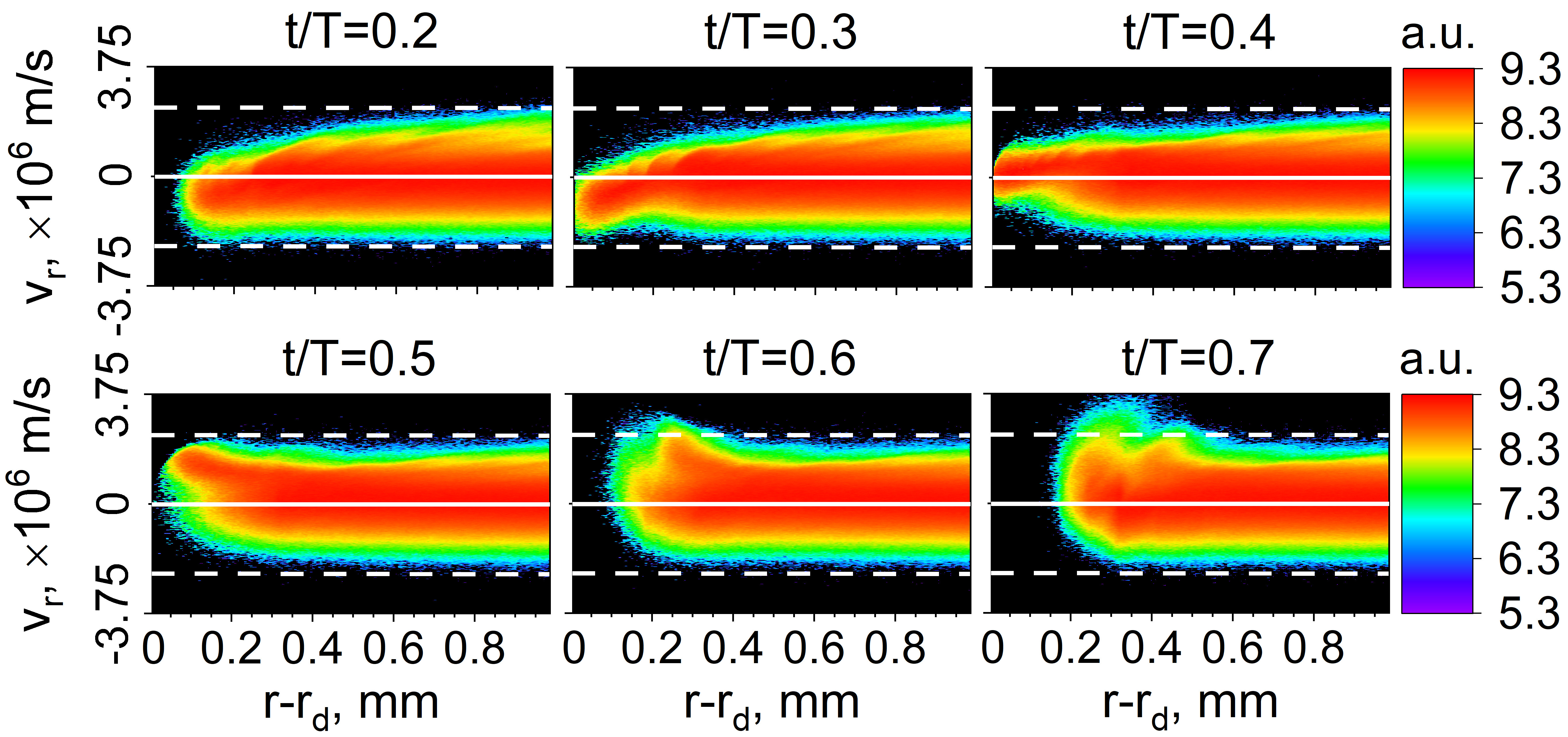}\label{fig5}
\caption{\label{fig5} Evolution of the decimal logarithm of the electron distribution function (measured in s/m$^4$) in the radial phase space around the sheath collapse and the start of sheath expansion. Solid white line indicates zero radial velocity, and dashed white lines indicate positive and negative radial velocities corresponding to the kinetic energy of radial motion equal to the argon ionization energy of $15.8$ eV.}
\end{figure}


We further argue that the observed generation of energetic electron beams is not caused by the plasma resonance electron heating. One can estimate the resonance width as \cite{aliev_1992} $\Delta = \nu_{\rm eff} L/\omega$ with $L=n_c/(dn_e/dr)$, $n_c$ being the cut-off density, and the effective frequency being determined either by electron-neutral elastic collisions or by the field convection caused by electron motion, $\nu_{\rm eff} = {\rm max}(\nu_{en},\omega({\rm v}_{Te}/\omega L)^{2/3})$. For the considered case with $10$ Pa and $1$ GHz shown in Fig.~\ref{fig4}, the effective frequency is determined by the second term, and the resonance width evaluates to $\Delta \approx 60 \mu$m, which should be resolved numerically as one can see features on that scale in Fig.~\ref{fig4}. While previous analyses of the plasma resonance heating (e.g., \cite{aliev_1993,ganachev_2004}) did not take into account the influence of electric field generated by the space charge in the sheath, simulations of the present work treat the electric field of both the sheath and the plasma resonance self-consistently. According to \cite{ganachev_2003,ganachev_2004}, the electric field at the plasma resonance should accelerate electrons in the direction of the dielectric boundary, hence it should be positive. The only candidate that appears close to the plasma resonance location is the reversed electric field at $r-r_d = 0.2$ mm during $t/T=(0.2,0.4)$. However, the plasma resonance curve always lies within the large negative electric field of the sheath (note also that it is also a region of a large positive space charge, which excludes excitation of plasma oscillations there since the latter require quasi-neutrality)
, and the positive electric field occurs only at a distance of $2\Delta \div 3 \Delta$ from the resonance location. Estimating the positive electric field to be around $10^4$ V/m and the corresponding region to be around $0.1$ mm wide, the corresponding electron energy gain is only about $1$ eV, and is thus insignificant. Moreover, the electrons energized by this mechanism immediately loose their energy by further interacting with the electric field of collapsing sheath. It follows that even if the observed reversed electric field could be ascribed to the plasma resonance, taking into account the electric field of the sheath makes its contribution to the net electron energization, which leads to the formation of the energetic electron beams seen in Fig.\ref{fig4}(d), negligible. Another argument in favor of the weak role of plasma resonance is that Fig.\ref{fig4}(c) exhibits two energetic electron beams, with the beams produced far from the plasma resonance location, especially for the second beam. Instead, such a pattern is common within CCRF discharges experiencing plasma series resonance \cite{Annaratone1995,Mussenbrock2008,wilczek_2016}, which can be regarded as a resonance with a normal mode becoming a surface mode for a sufficiently large plasma density \cite{eremin_2017a}. Therefore, it also corroborates the assumption that the electron energization results from interaction with the moving sheath, whose behavior is affected by interaction with the surface mode \cite{wilczek_2016,Eremin2023}. In particular, this leads to the observed asymmetry of the sheath behavior, since the expansion velocity is about two times higher than the collapse velocity.

\begin{figure}[t]
\centering
\includegraphics[width=6cm]{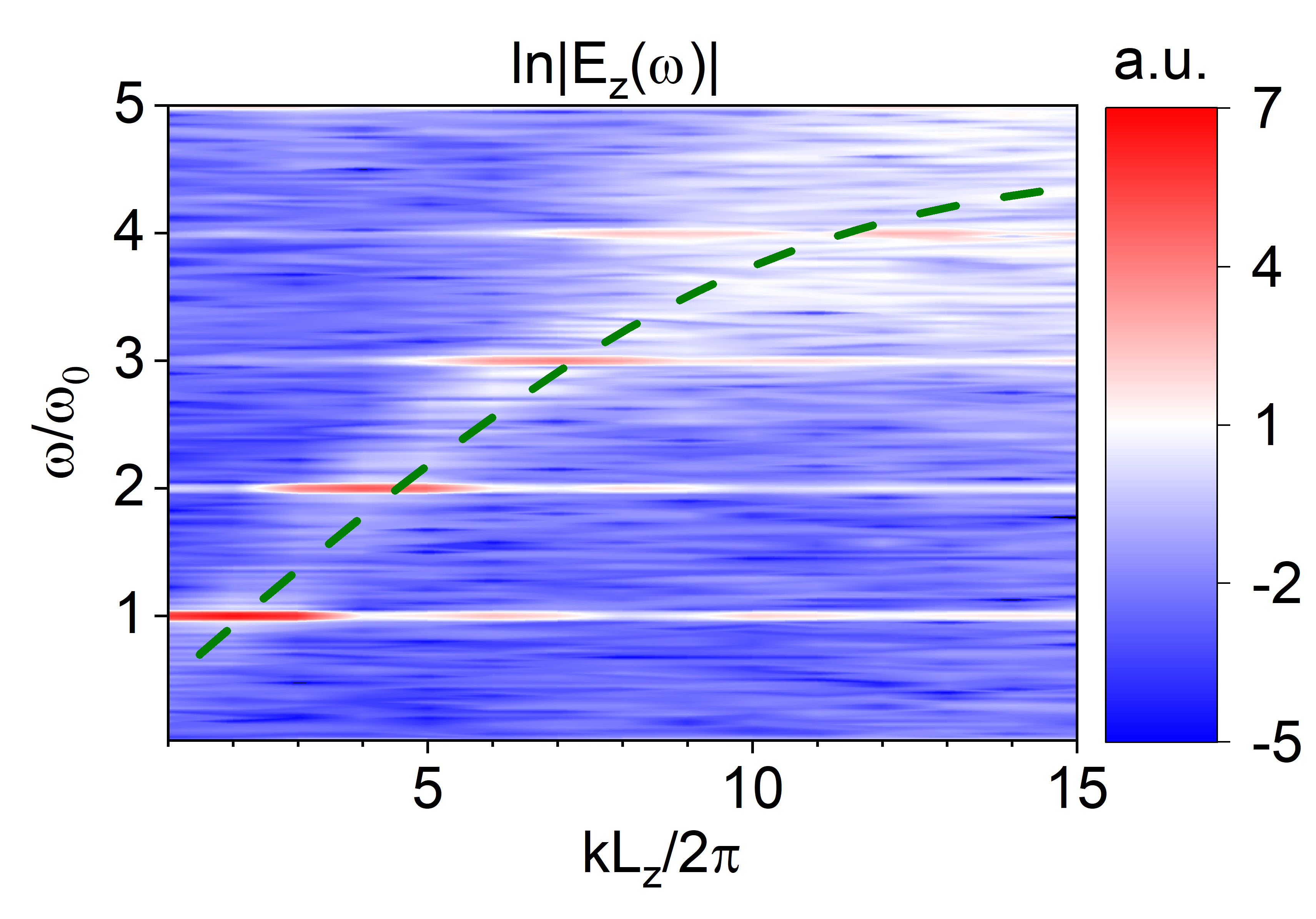}
\caption{\label{fig6} Natural logarithm of the Fourier axial electric field amplitude versus axial wavenumber plotted against the analytical dispersion relation solved from Eq. \ref{eq3} (dashed green curve).}
\end{figure}

To verify that there is significant excitation of the surface mode, we analyzed Fourier amplitude of the axial electric field components in the 2D ($\omega,k_z$) domain, see Fig.\ref{fig6}. It is instructive to compare this data with the analytical dispersion curve, that can be calculated as follows.
Neglecting collisions, and assuming harmonic spatiotemporal dependence $\propto e^{i(\omega t - kz)}$ for the field components,
the dispersion relation for the considered surface modes can be obtained from the equation
\begin{equation} \label{eq3}
\frac{\partial}{\partial r}\frac{1}{r\varepsilon_r}\frac{\partial}{\partial r}(rB_\theta)+\frac{\partial}{\partial z}\frac{1}{\varepsilon_r}\frac{\partial}{\partial z}B_\theta + k_0^2 B_\theta = 0.
\end{equation}
Assuming piecewise constant relative dielectric permittivities $\varepsilon_r = \{ \varepsilon_d, 1, \varepsilon_p = 1 - \omega_{pe}^2/\omega^2 \}$ for the dielectric (d), plasma sheath (ps), and plasma regions, continuity of $B_\theta$ and $E_z = -i c^2(\omega \varepsilon_r r)^{-1} \partial (rB_\theta)/\partial r$ at the boundaries between these regions, and the boundary condition $E_z = 0$ at the conducting surfaces of the metal rod (mr) and the outer wall (ow). In Eq.(\ref{eq3}) $k_0 = \omega/c$,
with $r_{mr}=2.4$ mm, $r_d=1.1$ cm, $\varepsilon_d = 3.8$, and $r_{ow}=4$ cm, and
the average parameters obtained from the PIC simulation of the $1$ GHz case $r_{ps}=1.14$ cm and $n_e=10^{18}$ m$^{-3}$. Solving this system, the dispersion relationship is shown by the dashed green line in Fig.\ref{fig6}. By comparing the latter with the PIC results, one can see that the azimuthal electric field Fourier amplitude exhibits maxima at the driving frequency harmonics, which are generated due to the nonlinear sheath behavior \cite{Lieberman2015}, and values of the azimuthal wavenumber are slightly trailing the analytical expectation, which is similarly observed in CCRF discharges \cite{eremin_2017,Eremin2023}. The Fourier amplitude is largest for the fundamental frequency, whereas the strongest excitation at this frequency occurs for $kL_{\rm z}/2\pi = 2$, which corresponds to the wavelength $\lambda = L_{\rm z}/2$. Since the distance between adjacent nodes of an SW is equal to half of the wavelength, one can expect four nodes for that SW, which can indeed be seen in Fig.~\ref{fig2}(d). Another strong excitation occurs at the second harmonic of the fundamental frequency. This can be related that to the asymmetry in the sheath behavior observed in Fig.~\ref{fig4}(a), where it is seen that the sheath expansion lasts about a quarter of a period, whereas the collapse phase is significantly slower and takes approximately half of the driving period.

Consequently, we have demonstrated the dominant role of a moving sheath in heating electrons within microwave surface wave discharges at low pressures, which bears many similarities to the physics of capacitively coupled discharges driven at radio frequencies. Simulations showed generation of energetic electron beams propagating towards the plasma bulk, as observed in previous experiments.
In contrast to some theoretical expectations, we have not found any evidence of the previously suggested plasma resonance heating. The conclusions drawn here can be applied to any surface wave discharge featuring a radial electric field perpendicular to the surface along which it propagates, which is the case for the vast majority of such discharges used for technological purposes.

\section*{Acknowledgments}

The authors appreciate fruitful discussions with Sebastian Wilczek and Ralf Peter Brinkmann. The work was supported by the German Research Foundation within the framework of the Sonderforschungsbereich SFB-TR 87 and by a Cooperative Research and Development Agreement between Applied Materials Inc. and PPPL, under contract number DE-AC02-09CH11466. We also thank Shahid Rauf and Kallol Bera for motivating this work. \\

\section*{Data Availability}

The data that support the findings of this study are available from the corresponding author upon reasonable request.

\bibliography{references}
\end{document}